\documentclass[conference]{IEEEtran}
\IEEEoverridecommandlockouts

\usepackage{cite}
\usepackage{amsmath,amssymb,amsfonts}
\usepackage{algorithmic}
\usepackage{graphicx}
\usepackage{textcomp}
\usepackage{xcolor}
\usepackage{url}
\usepackage{booktabs}
\usepackage{multirow}
\usepackage{comment}
\usepackage{subcaption}
\usepackage{colortbl}
\usepackage{multirow}
\usepackage{hyperref}
\usepackage{listings}
\usepackage{cases}
\usepackage{color}
\usepackage{arydshln}
\usepackage[mathscr]{euscript}
\usepackage[algoruled, boxed, vlined, linesnumbered]{algorithm2e}
\usepackage{makecell}
\usepackage{rotating}
\usepackage{balance}

\def\BibTeX{{\rm B\kern-.05em{\sc i\kern-.025em b}\kern-.08em
    T\kern-.1667em\lower.7ex\hbox{E}\kern-.125emX}}

\begin{document}

\title{When Surface Form Changes Moderation Decisions: A Paired Study of Code-Mixed Workflow Instability}

\author{
\IEEEauthorblockN{{\normalfont\textbf{Suraj Babu Thimma Krishnaram,
Yibo Hu,
Karthikeyan Saravanan}}
}

\IEEEauthorblockA{
Illinois Institute of Technology\\
Chicago, IL, USA\\
sthimmakrishnaram@hawk.illinoistech.edu, yhu89@illinoistech.edu,
ksaravanan1@hawk.illinoistech.edu
}
}

\maketitle

\begin{abstract}
Hate moderation is often evaluated as classification on clean English inputs, but deployed systems must route content to actions such as \textsc{Allow}, \textsc{Flag}, or \textsc{Review}. We study how this workflow changes under code-mixed inputs using a paired evaluation setting where the same underlying content is expressed as clean English and Tamil--English code-mix. Under thresholds tuned on clean English development data, code-mixed inputs produce substantial action instability, with a paired clean-to-code-mix decision flip rate of 0.265. The main workflow effects are increased review burden and increased false-flagging of non-hateful content: review rate rises from 0.138 to 0.297 and non-hate false-flag rate rises from 0.069 to 0.104. Tamil-only inputs show stronger degradation overall, suggesting a broader language-coverage limitation rather than the same code-mixed instability pattern. A simple disagreement-based deferral rule reduces automatic errors on stressed inputs, but only by increasing review load. These results show that workflow-level evaluation reveals moderation failures that standard classification summaries can miss.
\end{abstract}


\section{Introduction}

Most hate speech research evaluates models as classifiers on relatively clean inputs \cite{davidson-etal-2017-hateoffensive,founta-etal-2018-abusive,halim2023wokegpt}. Deployed moderation systems, however, must make operational decisions. A pipeline may automatically \textsc{Allow} a post, automatically \textsc{Flag} it, or defer it for human \textsc{Review}. This creates a mismatch between standard classification evaluation and the workflow behavior that matters in practice.

That mismatch becomes more important under realistic multilingual inputs. Social media users often write in code-mixed forms, combine words from multiple languages, or use romanized and non-standard spellings. Even when the underlying meaning is unchanged, these surface variations can change the final moderation action. Prior work has shown that multilingual and code-mixed settings expose robustness gaps that are easy to miss under standard benchmarking \cite{jin2023betteraskenglish,yang-chai-2025-codemixbench,winata2026can}. What remains less studied is how such variation changes routed moderation actions under a fixed operating point.

We study this question in a paired evaluation setting. Starting from labeled English hate-speech examples, we construct semantically aligned Tamil--English code-mixed variants and evaluate the same underlying content under multiple surface forms. This design lets us measure workflow-level instability directly, rather than comparing unrelated datasets. Our main interest is not whether code-mixed inputs uniformly increase all error types, but whether they change moderation behavior in operationally meaningful ways.

\begin{figure}[t]
\includegraphics[width=\linewidth, trim=100 100 100  100, clip]{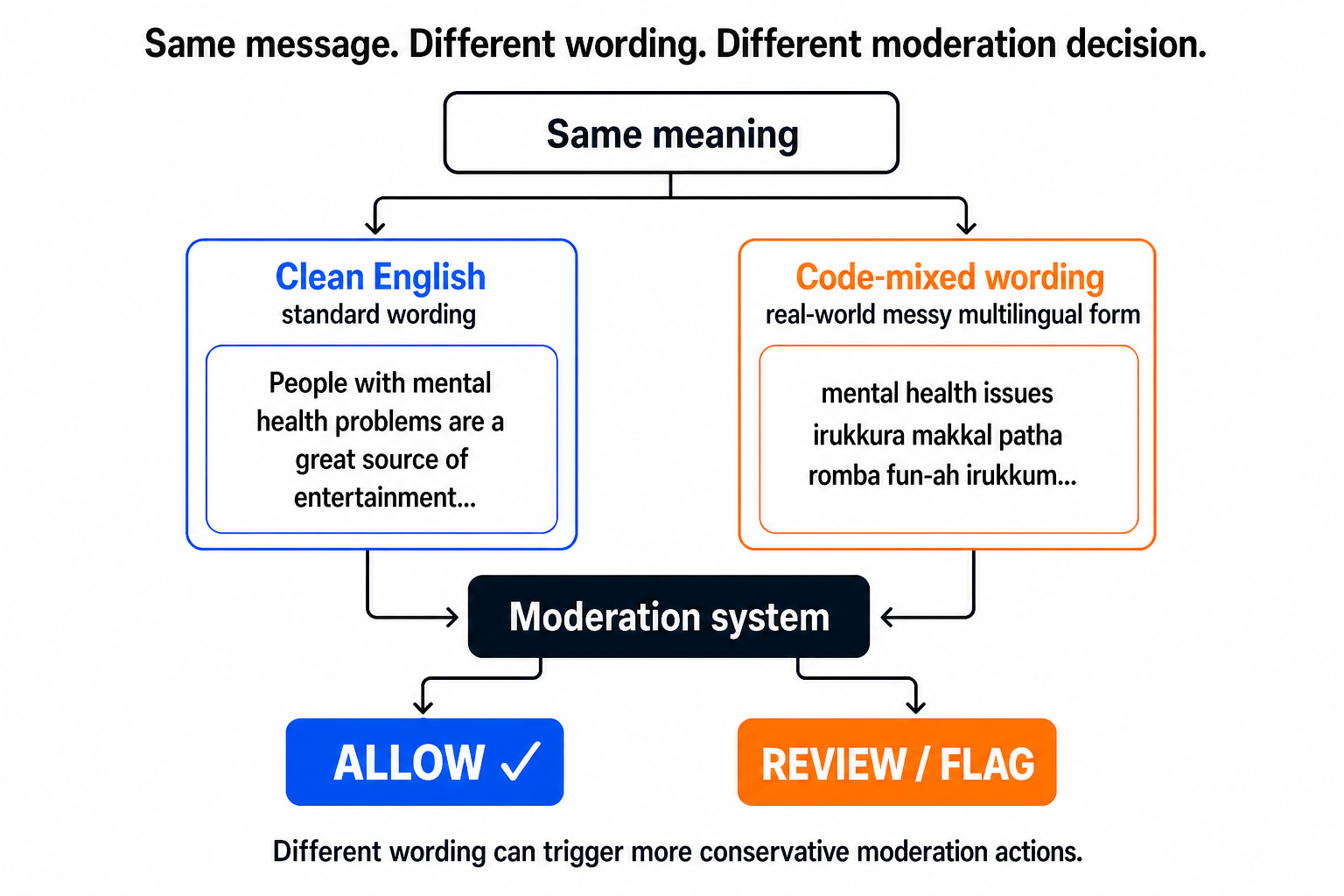}
\caption{A simple example of workflow instability.
A moderation system may allow a clean English post to pass automatically.
However, when the same underlying content appears in a real-world code-mixed form, surface variation can change the routed action, sending it to human review or flagging it as potentially harmful.
Same content, different wording, different moderation action.}
\label{fig:teaser}
\end{figure}

Our results show that they do. Under thresholds tuned on clean English development data, code-mixed inputs produce a paired clean-to-code-mix decision flip rate of 0.265. Relative to clean English, they also increase review rate from 0.138 to 0.297 and non-hate false-flag rate from 0.069 to 0.104. In other words, the main effect in our setting is not stronger false acceptance of hateful content, but instability that shifts more cases into review and incorrectly flags more non-hateful content.

To contextualize this pattern, we also evaluate Tamil-only inputs as a supporting diagnostic condition. Tamil-only degradation is stronger overall, suggesting a broader language-coverage limitation rather than the same workflow phenomenon as code-mixed instability. This helps separate two related but distinct effects: instability under mixed-language surface variation and degradation under full cross-lingual transfer.

Finally, we test a simple training-free disagreement-based deferral rule. When paired views receive different routed actions, we send the case to \textsc{Review}. This reduces automatic errors on stressed inputs, but only by increasing review load. Taken together, these results show that workflow-level evaluation reveals moderation failures that standard label-level summaries can miss.

Our contributions are:
\begin{itemize}
\item We study code-mixed hate moderation as a workflow-level action-routing problem centered on \textsc{Allow}/\textsc{Flag}/\textsc{Review} decisions rather than classification accuracy alone.
\item Using paired clean and Tamil--English code-mixed inputs, we show that code-mixed surface variation causes substantial moderation-action instability under fixed thresholds tuned on clean English development data.
\item We identify the main workflow effects of code-mix in our setting as increased review burden and increased non-hate false-flagging, rather than a uniform rise in all automatic errors.
\item We show that a simple disagreement-based deferral rule reduces automatic errors on stressed inputs, but only by increasing review load.
\end{itemize}

\noindent\textbf{Code availability.} Code and supplementary artifacts, including preprocessing, training, and evaluation notebooks, are available at \url{https://github.com/Surajtk/workflow-moderation-code#}.

\begin{figure*}[t]
\centering
\includegraphics[width=0.65\textwidth]{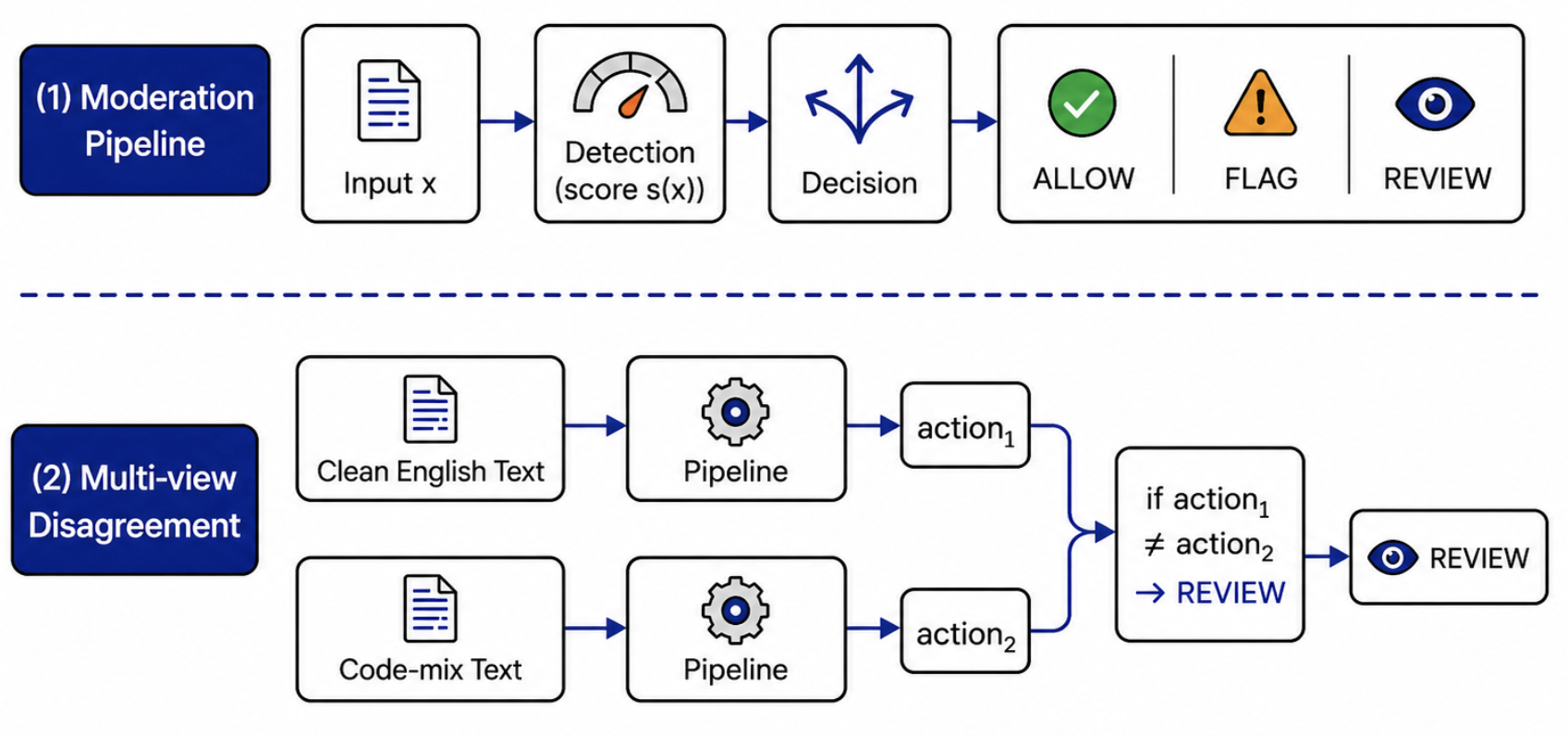}
\caption{Framework for workflow-level moderation and disagreement-based deferral.
A moderation pipeline routes each input to \textsc{Allow}, \textsc{Flag}, or \textsc{Review}.
For paired clean-English and code-mixed views of the same underlying content, action disagreement is treated as instability and routes the case to \textsc{Review}.}
\label{fig:pipeline}
\end{figure*}

\section{Related Work}

\subsection{Hate Speech Detection and Moderation}

A large body of work on hate speech, abusive language, and toxic content detection relies on labeled social media datasets and benchmark-style supervised evaluation \cite{davidson-etal-2017-hateoffensive,founta-etal-2018-abusive,halim2023wokegpt}. These studies established widely used classification baselines and helped define core prediction tasks for harmful language detection. However, they are typically evaluated with label-level metrics such as accuracy, F1, or AUC rather than the downstream moderation decisions that matter in deployment.

This distinction matters because moderation systems rarely operate as pure binary classifiers. In practice, platforms and safety pipelines must decide whether to allow content, automatically flag it, remove it, or route it for human review, often under asymmetric error costs and limited reviewer capacity. A model that appears strong under standard classification metrics may still behave poorly once its outputs are mapped into operational actions. Our work builds on the hate speech detection literature but shifts attention from label prediction alone to routed moderation behavior under a fixed operating point.

\subsection{Multilingual Robustness and Code-Mixed Stress Testing}

A separate line of work studies multilingual robustness, cross-lingual transfer, and code-mixed generation for hate and toxicity detection \cite{jin2023betteraskenglish,yang-chai-2025-codemixbench,winata2026can}. These resources show that surface-form variation can expose failures that remain hidden under standard English-only evaluation, including degradation under code-mixed and cross-lingual inputs. More broadly, multilingual hate and offensive language detection remains challenging because language coverage, script variation, domain mismatch, and uneven annotation quality can all weaken model reliability in realistic multilingual settings \cite{multilingual-hate-survey,multilingual-offensive-survey}.

Recent work also studies hate and offensive language detection directly in code-mixed settings, including both word-level and post-level tasks across different language pairs \cite{codemixed-hate-wordlevel,codemixed-hate-framework}. This literature reflects a practical reality rather than only a benchmark convenience: multilingual users often mix languages, scripts, and spelling conventions within the same utterance, especially in informal online contexts. As a result, semantically similar content can become harder to classify once it is expressed through mixed-language surface forms, non-standard spelling, transliteration, or script variation.

Our work is closest to this multilingual robustness literature, but it asks a different question. Prior benchmarks mainly study classification robustness and accuracy degradation under multilingual, code-mixed, or rewritten inputs. By contrast, we study whether paired clean and code-mixed versions of the same underlying content trigger different moderation actions under a fixed operating point. This moves the focus from predictive degradation alone to workflow instability in downstream moderation decisions.

Work on semantic-preserving adversarial perturbations makes a related point from a robustness perspective: relatively small surface changes can alter model behavior without substantially changing underlying meaning \cite{semantic-attack,semantic-attack-rl}. Our setting is not framed as adversarial attack generation, but it is motivated by the same core concern. Surface-form changes that preserve intended meaning can still reveal instability that standard benchmark evaluation misses, especially when those predictions are embedded in action-oriented moderation pipelines.

\subsection{Uncertainty, Abstention, and Human Review}

A related line of work studies uncertainty-aware prediction, abstention, and selective prediction in machine learning and NLP. In these settings, a model is allowed to abstain on low-confidence cases rather than being forced to make an automatic prediction, which can reduce errors on difficult or shifted inputs when the uncertainty signal is informative \cite{xin-etal-2021-art}. Prior work in NLP further shows that the usefulness of selective prediction depends heavily on the quality of confidence estimation and calibration \cite{xin-etal-2021-art,ulmer2022uncertainty}. More broadly, calibration methods aim to make model confidence better reflect empirical correctness, which is important whenever confidence scores are used to trigger abstention, escalation, or downstream decision rules \cite{guo2017calibration,kull2019dirichlet,hu2021multidimensional,hu2021uncertainty}.

This literature is closely related to moderation because abstention is often operationalized as deferral to human review. At the same time, deferral is not automatically neutral: who gets escalated, under what conditions, and with what error profile can create new fairness, workload, and resource-allocation trade-offs \cite{schreuder-chzhen-2021-classification}. In our setting, disagreement-based deferral is used as a deliberately simple baseline and analysis tool rather than proposed as a new abstention algorithm. We use paired multilingual views to show that workflow instability itself can act as a practical uncertainty signal under code-mixed variation.

Unlike prior robustness benchmarks, our focus is therefore on workflow-level moderation behavior under paired code-mixed stress, with particular emphasis on routed moderation actions such as \textsc{Allow}, \textsc{Review}, and \textsc{Flag}. This framing connects multilingual robustness to selective decision-making in moderation systems, where the key issue is not only whether a label changes, but whether the downstream action changes.
\section{Method}
\label{sec:method}

\subsection{Reference Moderation Pipeline}

We consider a reference moderation pipeline in which a pretrained detector is followed by a three-way decision layer: \textsc{Allow}, \textsc{Flag}, or \textsc{Review}. This abstraction isolates the operational choice that determines deployment risk---whether the system acts automatically or defers the case for review---while remaining simple enough to attribute changes in risk to input variation rather than additional pipeline complexity.

Given an input post $x$, the detector produces a hate-related risk score $s(x)$. A decision layer then maps the score to one of three actions:
\[
\begin{aligned}
\textsc{Allow}  &\text{ if } s(x) < \tau_{\mathrm{low}}, \\
\textsc{Flag}   &\text{ if } s(x) > \tau_{\mathrm{high}}, \\
\textsc{Review} &\text{ otherwise}.
\end{aligned}
\]
This triage formulation reflects practical moderation workflows and lets us study how surface-form variation changes routed actions without changing the underlying detector.

\subsection{Paired Stress Construction}

Our primary setting is paired clean versus code-mixed moderation. Starting from labeled English hate-speech examples, we construct semantically aligned Tamil--English code-mixed variants that introduce multilingual surface variation, including mixed-language tokens, romanized forms, and non-standard spellings, while preserving meaning as closely as possible.

The key design choice is pairing: for each source example, we evaluate the same underlying content under multiple surface forms rather than comparing unrelated datasets. We preserve the original hate/non-hate labels and use paired examples
\begin{itemize}
\item $x_{\mathrm{en}}$: the original English input,
\item $x_{\mathrm{mix}}$: a Tamil--English code-mixed variant,
\item $y \in \{\mathrm{hate}, \mathrm{non\mbox{-}hate}\}$: the inherited gold label.
\end{itemize}
This design lets us measure moderation-action changes for the same content under surface-form variation.

We focus on Tamil--English code-mixing as a controlled first testbed rather than a representative account of multilingual moderation. Code-mix generation follows the public CodeMixBench-style preprocessing pipeline \cite{shen-etal-2025-hatebench,yang-chai-2025-codemixbench}. We manually inspect a random subset of generated pairs for semantic preservation and label consistency, remove examples with clear semantic drift, malformed generation, or label inconsistency, and summarize the filtering procedure and validation sample size in the main text. Because we use a single generation pipeline, the results should be interpreted as specific to this construction process rather than universal across all code-mixed text.

To help interpret the main code-mix results, we also construct a Tamil-only comparison view for the same underlying test examples. This condition is diagnostic rather than primary: it helps distinguish code-mixed instability from broader language-coverage degradation under full cross-lingual transfer.

We use HateBenchSet as the source benchmark and follow the public CodeMixBench implementation as the base framework for generation and preprocessing.\footnote{\url{https://github.com/Jeromeyluck/CodeMixBench}} \footnote{\url{https://huggingface.co/datasets/TrustAIRLab/HateBenchSet}} HateBenchSet contains 7,838 manually labeled samples spanning 34 identity groups \cite{shen-etal-2025-hatebench}. We adapt this setup to construct paired English, Tamil--English code-mixed, and Tamil-only views for moderation evaluation.

To verify semantic preservation and label consistency, we manually inspect 500 randomly sampled generated pairs and remove low-quality or label-inconsistent examples before final evaluation. In total, we filter approximately 1,000 candidate rows (about 25\% of the initially generated set). After filtering, we retain paired test views of equal size for English, code-mix, and Tamil evaluation. The final evaluation uses a 738-example clean English development split for threshold selection and 3,017 paired test examples for each of the English, code-mixed, and Tamil views.

\subsection{Workflow-Level Risk Metrics}

Because our focus is moderation as action selection rather than label prediction alone, we use workflow-level metrics. Let $a \in \{\textsc{Allow}, \textsc{Flag}, \textsc{Review}\}$ denote the routed action and $y \in \{\mathrm{hate}, \mathrm{non\mbox{-}hate}\}$ the gold label. We report:
\begin{itemize}
\item \textbf{Hate False-Accept (HFA)}: $\Pr[a = \textsc{Allow} \mid y = \mathrm{hate}]$
\item \textbf{Non-Hate False-Flag (NFF)}: $\Pr[a = \textsc{Flag} \mid y = \mathrm{non\mbox{-}hate}]$
\item \textbf{Review Rate (RR)}: $\Pr[a = \textsc{Review}]$
\end{itemize}

HFA captures hateful content that is mistakenly allowed, NFF captures non-hateful content that is incorrectly flagged, and RR captures review workload. We also report paired action-instability statistics, especially the decision flip rate between clean English and code-mixed views.

\subsection{Operating Criterion}

We select an operating point by minimizing review rate subject to bounds on the two automatic-error metrics:
\[
\min_{\tau_{\mathrm{low}}, \tau_{\mathrm{high}}} \ \mathrm{RR}
\quad
\text{s.t.}
\quad
\mathrm{HFA} \le \epsilon_H,\;
\mathrm{NFF} \le \epsilon_{NH}.
\]

Here $\epsilon_H$ and $\epsilon_{NH}$ denote the target budgets for hate false-accept and non-hate false-flag. We treat false acceptance of hateful content as the more serious error and therefore use a stricter budget for HFA than for NFF. Unless otherwise noted, we use $\epsilon_H = 0.05$ and $\epsilon_{NH} = 0.10$.

Among all threshold pairs satisfying these constraints, we choose the pair with the lowest review rate. The selected thresholds are tuned on clean English development data only, then frozen and reused for clean, code-mixed, and Tamil test evaluations. In the main run, this procedure selects $\tau_{\mathrm{low}} = 0.031$ and $\tau_{\mathrm{high}} = 0.80$.

\subsection{Confidence-Based Abstention Baseline}

As a single-view uncertainty baseline, we evaluate confidence-based abstention using the detector outputs from the reference pipeline. Let $p_{\mathrm{hate}}(x)$ and $p_{\mathrm{non\mbox{-}hate}}(x)$ denote predicted class probabilities. We define confidence as
\[
c(x) = \max \bigl(p_{\mathrm{hate}}(x),\, p_{\mathrm{non\mbox{-}hate}}(x)\bigr).
\]

This baseline replaces the two-threshold routing rule with a single confidence threshold $\tau_{\mathrm{conf}}$. If $c(x) < \tau_{\mathrm{conf}}$, the input is routed to \textsc{Review}; otherwise, the predicted label is mapped directly to \textsc{Allow} or \textsc{Flag}. We tune $\tau_{\mathrm{conf}}$ on the clean English development split using the same risk-budget criterion and then freeze it for all test conditions. In the main run, this procedure selects $\tau_{\mathrm{conf}} = 0.965$.

This provides a direct comparison to disagreement-based deferral: confidence abstention uses a single-view internal score, whereas disagreement uses cross-view action disagreement as an external uncertainty signal.

\subsection{Disagreement-Based Deferral}

As a paired-view alternative to single-view confidence abstention, we evaluate a training-free disagreement rule. For each input, we construct two semantically related views---the original English input and an alternate view such as a code-mixed variant---and run the same moderation pipeline on both. If the routed actions agree, we keep that action; if they disagree, we route the case to \textsc{Review}. This treats action instability under semantically equivalent surface variation as a proxy for uncertainty.

We intentionally keep this mechanism simple. Our goal is not to optimize a production auxiliary-view pipeline, but to test whether paired-view disagreement provides useful workflow-level uncertainty signal under multilingual surface variation. In practice, such an auxiliary view could be produced by translation, transliteration, or multilingual normalization, each with distinct latency, compute, and failure-mode trade-offs. We do not implement or cost these upstream mechanisms here. Instead, we isolate the downstream question of whether cross-view action disagreement provides useful information once a semantically related alternate view is available.

\section{Experiments}

\subsection{Experimental Setup}

Our main detector is BERT-based multilingual cased\footnote{\url{https://huggingface.co/google-bert/bert-base-multilingual-cased}} fine-tuned for binary hate / non-hate classification on the English training view. We split the aligned dataset at the example level and derive paired English, code-mixed, and Tamil views from the same underlying examples. Thresholds are tuned on a clean English development split and then frozen for all downstream evaluations.

The main experiment evaluates the trained detector on paired English, code-mixed, and Tamil test views under the routing framework defined in Section~\ref{sec:method}. In addition to the reference two-threshold routing, we evaluate a confidence-based abstention baseline that routes low-confidence cases to \textsc{Review}, and a paired-view disagreement-based deferral variant that routes cross-view disagreements to \textsc{Review}. As supporting analyses, we also evaluate zero-shot HateBERT and zero-shot MetaHateBERT under the same routing framework.


The main detector is fine-tuned with maximum sequence length 320, learning rate 1.5e-5, batch sizes 8 (train) and 16 (eval), weight decay 0.01, warm-up ratio 0.1 and up to 4 epochs with early stopping patience 2. We select the best checkpoint by macro-F1 on the clean English development split. Thresholds for the downstream routing layer are tuned separately on the clean English development scores using the risk-budget criterion described in Section~\ref{sec:method}. The selected operating point for the main run is $\tau_{\mathrm{low}}=0.031$ and $\tau_{\mathrm{high}}=0.800$.

For disagreement-based deferral, we use paired clean/alternate views, specifically clean--code-mix and clean--Tamil pairs. Disagreement between the routed actions of the two views is treated as an uncertainty signal, and such cases are routed to \textsc{Review}. For confidence-based abstention, we use the detector's maximum predicted class probability as a single-view confidence score and defer cases whose confidence falls below a tuned threshold.

\subsection{Threshold Selection}

We select $(\tau_{\mathrm{low}}, \tau_{\mathrm{high}})$ on a clean development set and keep them fixed for all downstream test conditions, including code-mixed evaluation. This choice matches our robustness setting: the deployment operating point is determined under standard inputs, and we then measure how workflow-level risk changes under code-mixed variation.

In practice, we sweep threshold pairs on the clean English development split, compute HFA, NFF, and RR for each pair, and choose the operating point that satisfies the risk budget in Section~\ref{sec:method} with the lowest review rate. We sweep $\tau_{\mathrm{low}}$ over $\{0.001, 0.006, \dots, 0.196\}$ and $\tau_{\mathrm{high}}$ over $\{0.800, 0.805, \dots, 0.995\}$, considering only pairs with $\tau_{\mathrm{low}} < \tau_{\mathrm{high}}$.

The thresholds are tuned on the 738-sample clean English development split using the risk-budget criterion in Section~\ref{sec:method}, with $\epsilon_H = 0.05$ and $\epsilon_{NH} = 0.10$. The selected operating point for the main mBERT system is $\tau_{\mathrm{low}} = 0.031$ and $\tau_{\mathrm{high}} = 0.800$, which achieves HFA $= 0.0478$, NFF $= 0.0733$, and RR $= 0.1341$ on the development split.

Unless otherwise noted, all evaluated routing variants use clean-dev-tuned operating points without retuning on stressed conditions. For the reference three-action baseline and the disagreement-based deferral variants, we use the clean-dev-tuned thresholds $\tau_{\mathrm{low}} = 0.031$ and $\tau_{\mathrm{high}} = 0.800$. For the confidence-based abstention baseline, we tune a single confidence threshold on the same clean development split under the same risk-budget criterion and obtain $\tau_{\mathrm{conf}} = 0.965$. This ensures that any downstream change reflects robustness differences rather than threshold adaptation to the test setting.

\subsection{Main Comparisons}

Our main comparisons are:
\begin{itemize}
\item clean, code-mixed, and Tamil baselines,
\item clean, code-mixed, and Tamil with confidence-based abstention,
\item code-mixed and Tamil with disagreement-based deferral.
\end{itemize}

These comparisons test three questions: (1) how stressed multilingual views change workflow-level moderation behavior under a fixed pipeline, (2) whether single-view confidence abstention improves the safety--workload trade-off, and (3) whether paired-view disagreement provides a more useful abstention signal than ordinary model confidence.

\subsection{Supporting Analyses}

We include supporting analyses of detector choice, paired-view instability across clean--code-mix and clean--Tamil comparisons, and sensitivity to the operating budget.
\section{Results}

\subsection{Main Findings}

Table~\ref{tab:main} shows that multilingual surface variation materially changes moderation behavior under fixed thresholds. Relative to clean English, code-mixed inputs increase review rate from 0.138 to 0.297 and increase non-hate false-flag rate from 0.069 to 0.104, while hate false-accept decreases from 0.040 to 0.011. The main effect in the current setup is therefore not a uniform rise in all automatic errors, but a redistribution of actions toward review and, for code-mix, increased false-flagging of non-hateful content.

\begin{table}[htbp]
\centering
\caption{Main workflow-level results. HFA = hate false-accept, NFF = non-hate false-flag, and RR = review rate. Lower HFA and NFF are better; higher RR indicates greater review burden.}
\begin{tabular}{lccc}
\toprule
Setting & HFA $\downarrow$ & NFF $\downarrow$ & RR $\uparrow$ \\
\midrule
Clean baseline & 0.040 & 0.069 & 0.138 \\
Code-mix baseline & 0.011 & 0.104 & 0.297 \\
Tamil baseline & 0.009 & 0.060 & 0.637 \\
Clean + confidence & 0.041 & 0.031 & 0.197 \\
Code-mix + confidence & 0.016 & 0.032 & 0.417 \\
Tamil + confidence & 0.012 & 0.007 & 0.786 \\
Code-mix + disagreement & 0.003 & 0.046 & 0.355 \\
Tamil + disagreement & 0.005 & 0.014 & 0.675 \\
\bottomrule
\end{tabular}
\label{tab:main}
\end{table}

Tamil produces a stronger overall shift in workflow behavior. Relative to clean English, Tamil increases review rate from 0.138 to 0.637, decreases hate false-accept from 0.040 to 0.009, and does not show the same concentrated non-hate false-flag increase as code-mix. This pattern suggests broader language-coverage degradation rather than the more specific code-mixed instability pattern seen in the clean-to-code-mix comparison.

\begin{table*}[t]
\centering
\small
\caption{Paired decision instability between clean and stressed views. Flip is the fraction of paired examples whose routed action changes across views; CIs are 95\% bootstrap confidence intervals. HFA = hate false-accept, NFF = non-hate false-flag, and RR = review rate. $\Delta$ values are stressed minus clean.}
\label{tab:flip}
\begin{tabular}{lccccc}
\toprule
Comparison & Flip & 95\% CI & $\Delta$HFA & $\Delta$NFF & $\Delta$RR \\
\midrule
Clean $\rightarrow$ code-mix & 0.266 & [0.251, 0.281] & -0.029 & 0.035 & 0.160 \\
Clean $\rightarrow$ Tamil    & 0.557 & [0.539, 0.573] & -0.031 & -0.009 & 0.500 \\
\bottomrule
\end{tabular}
\end{table*}

\begin{figure*}[htbp]
\centering
\includegraphics[width=0.9\linewidth]{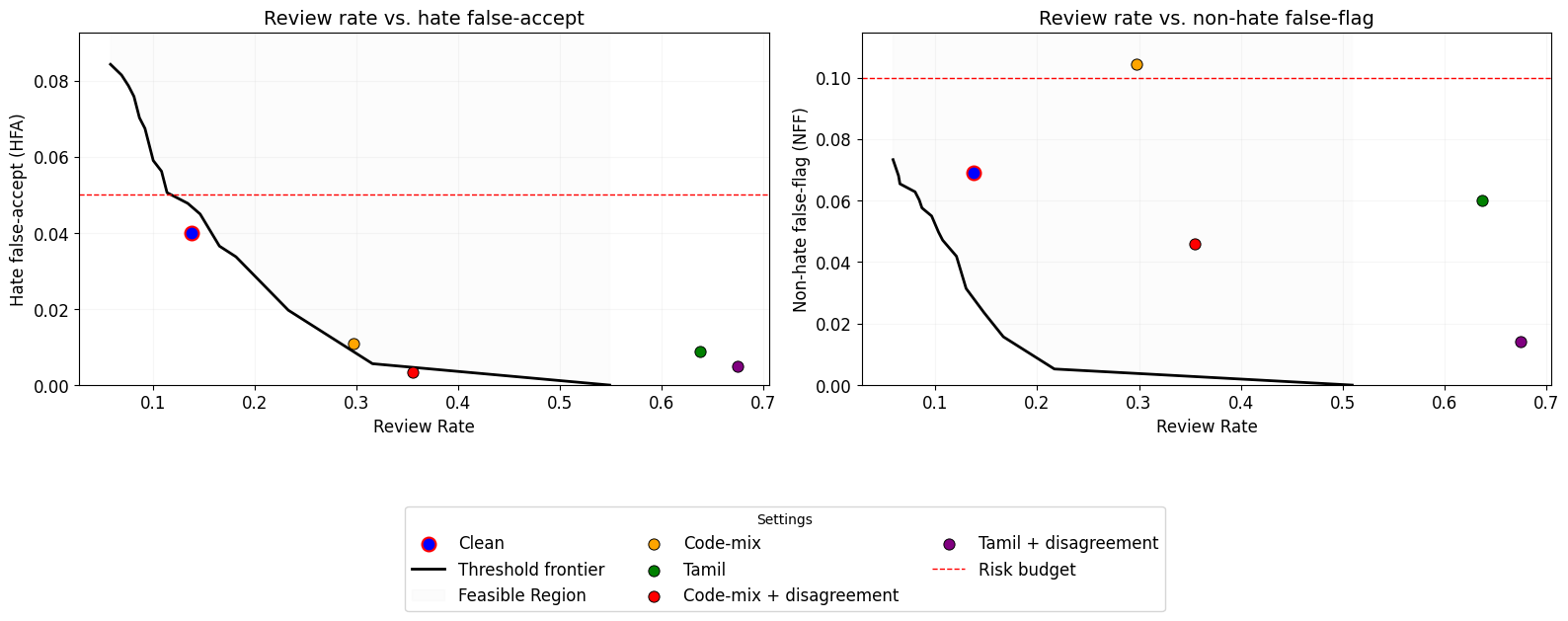}
\caption{Risk--review trade-offs under different threshold pairs, depicting the efficiency frontier and risk budgets. The left panel shows review rate vs. hate false-accept, and the right panel shows review rate vs. non-hate false-flag.
}
\label{fig:risk_review}
\end{figure*}

Confidence-based abstention provides a stronger single-view baseline than the reference two-threshold pipeline, but it does not dominate paired-view disagreement-based deferral. On code-mixed inputs, confidence abstention reduces non-hate false-flagging from 0.104 to 0.032, but increases review rate from 0.297 to 0.417 and yields hate false-accept 0.016. By contrast, disagreement-based deferral yields substantially lower hate false-accept at 0.003 with a lower review rate of 0.355, although its non-hate false-flag rate is higher at 0.046. The same trade-off appears more sharply for Tamil, where confidence abstention drives review rate to 0.786, while disagreement-based deferral achieves lower HFA and lower review burden. Figure~\ref{fig:risk_review} places the selected threshold pair within the broader threshold space and makes the operational trade-off explicit: lower automatic risk generally requires higher review cost.

Taken together, these results show that multilingual stress changes moderation primarily through action redistribution rather than through a uniform degradation in all error types. Code-mixing increases review burden and non-hate over-enforcement, whereas Tamil produces a broader shift toward review consistent with weaker language coverage. This is why workflow-level evaluation is necessary here: the same model can appear safer on one metric while becoming more operationally costly or less reliable overall.

\subsection{Paired Instability}

Paired evaluation confirms that the same underlying content often receives different routed actions under alternate surface forms. The clean-to-code-mix flip rate is substantial, and the clean-to-Tamil flip rate is larger still, showing that workflow instability is not a marginal effect. Because these comparisons are paired at the content level, the observed differences reflect changed moderation actions rather than differences in dataset composition.

McNemar tests show significant paired differences for clean versus code-mix on HFA, NFF, and RR, and for clean versus Tamil on HFA and RR, while the Tamil NFF difference is not significant. Together with the confidence intervals in Table~\ref{tab:flip}, these results support the central claim of the paper: moderation robustness should be evaluated at the workflow level, because surface-form variation can materially change downstream actions even when label-level summaries understate the effect.

\subsection{Label-Conditioned Transitions}

\begin{table*}[t]
\centering
\small
\setlength{\tabcolsep}{6pt}
\caption{Row-normalized clean-to-code-mix action transition matrices by label. Each row sums to 1.000. For non-hate examples, clean \textsc{Allow} and \textsc{Review} decisions move more often toward \textsc{Review} and \textsc{Flag}; for hate examples, clean \textsc{Allow} decisions move away from mistaken \textsc{Allow}.}
\label{tab:transition_by_label}
\begin{tabular}{lccc@{\hspace{1.5cm}}lccc}
\toprule
\multicolumn{4}{c}{Non-hate} & \multicolumn{4}{c}{Hate} \\
\cmidrule(r){1-4} \cmidrule(l){5-8}
Clean $\rightarrow$ Code-mix & \textsc{Allow} & \textsc{Review} & \textsc{Flag}
& Clean $\rightarrow$ Code-mix & \textsc{Allow} & \textsc{Review} & \textsc{Flag} \\
\midrule
\textsc{Allow}  & 0.646 & 0.339 & 0.015 & \textsc{Allow}  & 0.086 & 0.793 & 0.121 \\
\textsc{Review} & 0.019 & 0.698 & 0.283 & \textsc{Review} & 0.045 & 0.580 & 0.376 \\
\textsc{Flag}   & 0.009 & 0.324 & 0.667 & \textsc{Flag}   & 0.003 & 0.112 & 0.885 \\
\bottomrule
\end{tabular}
\end{table*}

Table~\ref{tab:transition_by_label} shows that code-mixed instability is operationally asymmetric. For non-hate examples, many clean \textsc{Allow} and \textsc{Review} decisions shift toward more conservative actions, consistent with increased review burden and non-hate false-flagging. For hate examples, many clean \textsc{Allow} decisions move to \textsc{Review} or \textsc{Flag}, helping explain why hate false-accept decreases under code-mixed rewriting. Tamil shows a broader coverage failure overall, whereas code-mix more specifically exposes instability under partial surface variation.

Table~\ref{tab:qual_examples} provides illustrative examples of the workflow-level action changes quantified above.

\begin{table*}[t]
\centering
\small
\setlength{\tabcolsep}{5pt}
\caption{Illustrative workflow-level action changes under code-mixed rewriting. Labels indicate gold annotations; arrows show routed action changes from clean to code-mixed inputs.}
\label{tab:qual_examples}
\begin{tabular}{
p{0.06\textwidth}
>{\raggedright\arraybackslash}p{0.22\textwidth}
>{\raggedright\arraybackslash}p{0.34\textwidth}
>{\raggedright\arraybackslash}p{0.34\textwidth}}
\toprule
Case & Change & Clean & Code-mix \\
\midrule
1 & Non-hate: \textsc{Allow}$\rightarrow$\textsc{Flag} &
Pacific Islanders are an indigenous racial group... &
``Pacific Islanders oru racial group... indigenous...'' \\
2 & Non-hate: \textsc{Review}$\rightarrow$\textsc{Flag} &
I don't think there is anything bad about being a Muslim... &
``Muslim aga iruppathil entha problem-um illa...'' \\
3 & Hate: \textsc{Allow}$\rightarrow$\textsc{Review} &
People with mental health problems are a great source of entertainment... &
``mental health issues irukkura makkal... fun-ah irukkum...'' \\
4 & Hate: \textsc{Allow}$\rightarrow$\textsc{Flag} &
Asians are cunning and lie a lot... &
``Asian-nga romba cunning-ah irukkanga...'' \\
\bottomrule
\end{tabular}
\end{table*}

These examples are consistent with the broader transition patterns rather than isolated anecdotes. In particular, non-hate examples illustrate escalation from \textsc{Allow} or \textsc{Review} to more conservative actions, while hate examples show movement away from mistaken \textsc{Allow} decisions. The examples therefore make the operational asymmetry in Table~\ref{tab:transition_by_label} concrete at the instance level.

The label-conditioned and qualitative views together clarify that code-mixed robustness failure is not a single phenomenon. For non-hate content, the dominant effect is over-escalation into review and flagging, whereas for hate content the more visible shift is away from mistaken \textsc{Allow} decisions. This asymmetry helps explain why code-mixed inputs can reduce hate false-accept while simultaneously increasing review burden and non-hate over-enforcement.

\subsection{Detector Sensitivity}
\label{app:detector_analysis}

Table~\ref{tab:detector_ablation} shows that code-mixed instability varies across backbones. mBERT has a flip rate of 0.265, MetaHateBERT reaches 0.608 together with a sharp increase in non-hate false-flagging, and HateBERT collapses to a near all-review regime. Detector choice therefore changes how multilingual stress redistributes workflow risk, but does not remove the need for action-level evaluation.

This variability across backbones suggests that multilingual robustness is partly a model-selection problem, but not one that can be diagnosed from label accuracy alone.

\begin{table*}[t]
\centering
\small
\setlength{\tabcolsep}{5pt}
\caption{Detector-wise workflow results for the clean$\rightarrow$code-mix setting. Thresholds are tuned on clean English development data and frozen for test-time evaluation. HFA = hate false-accept, NFF = non-hate false-flag, and RR = review rate.}
\label{tab:detector_ablation}
\begin{tabular}{>{\raggedright\arraybackslash}p{0.27\textwidth}ccccccc}
\toprule
Model & Clean HFA & Code-mix HFA & Clean NFF & Code-mix NFF & Clean RR & Code-mix RR & Flip rate \\
\midrule
HateBERT (zero-shot) & 0.000 & 0.000 & 0.000 & 0.000 & 1.000 & 1.000 & 0.000 \\
MetaHateBERT (zero-shot) & 0.045 & 0.014 & 0.029 & 0.436 & 0.463 & 0.443 & 0.608 \\
mBERT (fine-tuned) & 0.040 & 0.011 & 0.069 & 0.104 & 0.138 & 0.297 & 0.265 \\
\bottomrule
\end{tabular}
\end{table*}

\section{Conclusion}

This paper studies code-mixed hate moderation as a workflow-level problem. Using paired clean English and Tamil--English code-mixed views of the same underlying content, we show that multilingual surface variation can materially change routed moderation decisions under fixed thresholds tuned on clean English development data. In our setting, code-mixed inputs increase review burden, increase false-flagging of non-hateful content, and produce substantial paired decision instability, while Tamil-only inputs show a stronger degradation pattern consistent with broader language-coverage failure.

We also show that a paired-view disagreement-based deferral rule reduces automatic errors on stressed inputs, but only by increasing review load. More broadly, the results distinguish two related robustness problems: code-mixed inputs expose workflow instability under partial surface variation, whereas Tamil-only inputs reflect a broader language-coverage failure. These findings suggest that moderation robustness should be evaluated at the level of deployed actions---\textsc{Allow}, \textsc{Review}, and \textsc{Flag}---because label-level summaries can understate operational risk under multilingual surface variation.

\section{Limitations}
This work is intentionally narrow in scope. We study a reference moderation pipeline in a single Tamil--English setting as a controlled testbed for decision-level robustness, so the results should not be interpreted as a comprehensive account of multilingual moderation across languages, platforms, or policy regimes. We evaluate robustness under a fixed pipeline rather than optimizing multilingual performance through retraining, calibration, language identification, translation-based preprocessing, or stronger multilingual baselines.

The stressed code-mixed views are automatically generated and manually filtered on a subset rather than exhaustively validated, so residual generation artifacts or imperfect semantic preservation may remain. Because approximately 25\% of initially generated candidates are filtered before final evaluation, the retained benchmark may underrepresent especially noisy, ambiguous, or hard-to-preserve examples. We therefore interpret the dataset as a controlled multilingual stress test rather than a direct estimate of naturally occurring code-mixed traffic.

Our disagreement-based deferral rule also assumes access to paired semantic views at inference time, but this paper does not implement or cost a production mechanism for generating such views, such as transliteration, translation, or multilingual normalization. As a result, we do not measure the latency, compute overhead, or failure modes of the auxiliary-view pipeline itself.

Finally, increased deferral is not operationally neutral. Higher review rates may concentrate additional scrutiny or delay on users and communities that code-mix more frequently. This paper quantifies the workflow trade-off but does not evaluate subgroup-specific impacts, reviewer-capacity constraints, or downstream fairness effects. We also do not test calibrated confidence baselines or a broader set of modern multilingual backbones, so our comparison between disagreement-based deferral and confidence-based abstention should be interpreted within the experimental scope studied here.
\section{Ethical Considerations}

This work studies hate moderation, a domain in which both under-enforcement and over-enforcement can cause harm. False accepts of harmful content may leave harmful material visible, while false flags and increased review rates may impose additional scrutiny, delay, or friction on users whose language practices differ from the clean-English setting used to tune the system. In particular, if code-mixing is more common in specific communities, a review-heavy mitigation strategy could shift moderation burden unevenly onto those groups. The qualitative examples are included only to illustrate workflow-level failure modes and are lightly abbreviated or romanized for readability. This work does not release a deployable moderation system; its goal is to expose robustness and workflow risks that should be monitored before deployment.

\section*{Acknowledgment}
This work used computational resources provided by the Chameleon testbed~\cite{keahey2020lessons}, which is supported by the National Science Foundation.

\balance
\bibliography{reference}
\bibliographystyle{IEEEtran}

\end{document}